\documentclass[runningheads]{llncs}
\usepackage{graphicx}
\usepackage{algorithm}

\begin{document}
\title{Performance Evaluation of Multiparty Authentication in 5G IIoT Environments}
\titlerunning{Evaluating Multiparty Authentication for 5G Networks}
\author{Hussain Al-Aqrabi\inst{1}\orcidID{0000-0003-1920-7418} \and
Phil Lane\inst{2,3}\orcidID{0000-0002-2179-0166} \and
Richard Hill\inst{3}\orcidID{0000-0003-0105-7730}}
\authorrunning{Al-Aqrabi et al.}
\institute{School of Computing and Engineering, University of Huddersfield, UK
\email{\{h.al-aqrabi,p.lane,r.hill\}@hud.ac.uk}}

\maketitle              % typeset the header of the contribution
\begin{abstract}
%The abstract should briefly summarize the contents of the paper in
%15--250 words.
With the rapid development of various emerging technologies such as the Industrial Internet of Things (IIoT), there is a need to secure communications between such devices.  Communication system delays are one of the factors that adversely affect the performance of an authentication system. 5G networks enable greater data throughput and lower latency, which presents new opportunities for the secure authentication of business transactions between IIoT devices. We evaluate an approach to developing a flexible and secure model for authenticating IIoT components in dynamic 5G environments. 
\end{abstract}
\section{Introduction}
Fifth generation (5G) networks are becoming more recognised as a significant driver of Industrial Internet of Things (IIoT) application growth \cite{kalra2015,alaqrabi2015}. Recent developments in the growth of wireless and networking technologies such as software-defined networking and hardware virtualisation have led to the next generation of wireless networks and smart devices. In comparison to 4G technologies, 5G is characterised by: higher bit rates (more than 10 gigabits per second), higher capacity, and very low latency \cite{ferrag2018}, which is a significant asset for the billions of connected devices in the context of Internet of Things (and Industrial Internet of Things) domains. As such, emerging IoT applications and business models require new approaches to measuring performance, utilising criteria such as security, trustworthy, massive connectivity, wireless communication coverage, and very low latency for a large number of IoT devices.
\\
The introduction of 5G infrastructure is particularly important for IoT and Industrial IoT (IIoT) as it supports increases in data throuput transmission rates, as well as reducing system latency.

As IoT devices proliferate and we delegate more tasks to them, there is a greater need to a) exchange data, and b) augment existing data transmission to facilitate improved trust mechanisms between objects. In any network, an increase in the size and volume of data packets that are transported can increase response times, which is undesirable in a highly interconnected environment of smart objects. 5G performance facilitates reduced response times between communicating entities, which helps enhance the user experience.

Together, these characteristics directly support the closer cooperation of decoupled physical objects, and is the basis for a more connected future.

5G deployments in the millimeter wave region of the radio spectrum is one of the main enabling factor for improved network performance, albeit at a loss of transmission distance. Operating at $\approx$ 30~GHz offers some physical security due to the limited propagation ranges available.  However, the industrial scenario of a malicious employee, stood beside some manufacturing equipment that is IIoT enabled, means that data can potentially be `sniffed' and relayed to an external system.

Awareness and accessibility of technologies that can connect the operations of devices, with a view to permitting innovative ways of optimising operations and minimising waste through cooperation, is a key objective of the Industry 4.0 movement. Industrial organisations have mission-critical intellectual property (IP) that is central to their ability to compete effectively and maintain profits now and into the future. The detail and insight that describes such IP is contained within the myriad industrial operations that take place, and therefore the use of IIoT, which generally includes wireless networking technologies, is a concern to many who wish to protect their IP.

The traditional methods of ensuring secure network communications tend to depend on the checking of credentials against a central authority. This has proven satisfactory in many cases, though this is up to a point as systems grow to the point where the authentication mechanism itself can become a bottleneck. If we consider the potential number of connected parties that would require authentication in an IIoT environment, it is clear that a centralised authentication system cannot scale sufficiently without harming the operation of the whole system.

%We therefore need to be able to create a scalable security architecture that can robustly marshall the suitable authentications for different parties, while accommodating the dynamic nature of how IIoT devices communicate at the same time.
IIoT devices are dynamic, often mobile, and need to work and be trusted for variable amounts of time, usually to complete a particular transaction in a timely fashion. The potential requirements of IIoT are such that any architecture that is essential for the secure exchange of data, must also be scalable to meet what seems to be an inconceivable future demand. 
\subsection{Industrial Internet of Things in the 5G Era}
The IIoT has the ability to provide intelligent services to users, while presenting privacy and security issues and perhaps new challenges to standards and governance bodies \cite{ferrag2018}. Research studies have focused on state-of-the-art research in several aspects of IIoT and 5G technologies, from academic and industrial perspectives \cite{ldxu2014,hosek2013}. The aim is to find a place for recent developments in theory, application, standardisation and the application of fifth-generation technologies in IIoT scenarios \cite{Ishaq2013}.

5G technology has the potential to expand IIoT capabilities, significantly beyond what is feasible with existing technology. A 5G wireless network will enable IIoT devices to communicate to a new level within smart environments, through connected `smart sensors'. In addition, a 5G wireless network may also considerably expand the scope and scale of IIoT coverage by offering the fastest communication and capacity for business transactions\cite{Akpakwu2017}.

Whilst IoT systems are often aimed at enhancing the quality of everyday life, including interconnecting users, smart home devices, and smart environments such as smart cities and smart homes, the IIoT is a domain of significant interest as it is through the enhanced coordination and optimisation possibilities of inter-connected manufacturing operations that industrial organisations can become more competitive.

However, IIoT adoption is still developing in industry and faces several challenges, including new demands for product and solutions, and the transformation of business models. In some industries, such as healthcare, or traffic management, etc., the IIoT still must overcome several technical challenges such as flexibility, reliability and robustness \cite{Elkhodr2017}.

5G-enabled IoT can make important contributions to the future of IIoT by connecting billions of IoT devices to generate a huge `network of things' in which intelligent devices interact and share data without any human assistance. Presently, a heterogeneous application domain makes it very hard for IIoT to identify whether the individual system components are capable of meeting application requirements \cite{alaqrabi2013}.

\subsection{Key challenges for 5G-IoT architecture }

While a lot of studies have been done on 5G IoT, there are still technical challenges to overcome. In this section, we will briefly review the major challenges for 5G IoT.

Due to the openness of network architectures and rapid communication network deployment of a wide variety of services, 5G IoT systems pose major challenges for information security, increased privacy concerns, trusted communications between devices, etc. Several researchers have contributed to strengthen authentication mechanisms  \cite{he2012,deng2009,kar2011} in 4G and 5G cellular networks and there are various cryptographic algorithms to address potential security and privacy problems for 4G and 5G networks \cite{Alrawais2017,barni2010,ma2014,mahmoud2016}. The emerging interest in microservices architecture\cite{hill2017} emphasises the need to consider how trust can be engendered between ever-decreasing units of computation.

Although there are numerous heterogeneous mechanisms for secure communication, 5G IoT integrates a number of different technologies and this has an important impact upon IoT applications \cite{bessis2013,hill2013}. 

As the amount of devices in IoT networks increases every day, the management of these devices is becoming increasingly complex\cite{kalra2015}. Due to the large amount of IoT devices, scalability of the network and network management is a significant problem in 5G-IoT. To manage the state information such a large number of devices IoT devices with satisfying performance is also a problem that needs to be addressed.  Also, several current IoT applications comprise of overlaid deployments of IoT devices networks where both applications and devices are unable to communicate and share information. 

These devices need to be able to flexibility connect the network at any time. Since IoT systems generate and/or process sensitive information, it must authenticate itself to obtain and deliver information to the gateway.

Furthermore, the capability and effectiveness to gather and distribute information in the physical globe is challenging. There are still several challenges remaining for 5G IoT that need to be addressed, such as the seamless interconnection between heterogeneous communication networks where a large amount of IoT heterogeneous devices are connected via a complex communication network with varying technology to communicate, and retrieve vital information with other intelligent networks or applications. High availability of IoT devices is essential for real-time monitoring systems, as they need to be accessible to monitor/collect data. IoT devices that may be compromised and vulnerable to hackers, physically harmed or stolen, resulting in service disruption, and it is not easy to locate an impacted node. 
\section{Dynamic multiparty authentication}
Due to the rapid development of various emerging technologies and computing paradigms, such as Mobile Computing and the Internet of Things presents significant security and privacy challenges \cite{sotiriadis2013,Elkhodr2017}.

With the explosive development of the Internet of Things applications, the shift from traditional communication facilities to the Internet is becoming increasingly crucial for group communication.

Many new Internet services and applications are emerging, such as cloud computing that allows users to elastically scale their applications, software platforms, and hardware infrastructure \cite{baker2012}. Cloud-based business systems are dynamic in a multi-tenancy setting and require likewise dynamic authentication relationships. 

Therefore, the authentication frameworks can not be static. However, experience in the domain of Cloud computing helps the comprehension of how IoT applications can be subject to security threats, such as exploiting virtue, malware attacks, distributed attacks, and other known cloud challenges \cite{alaqrabi2013}.

These cloud implementations increase the sharing of resources that can be made available by dividing solutions into distinct levels. Consequently, the increasing proliferation of services provided by IoT technologies also presents many security and privacy-related risks.

Within the shared domains of IoT cloud, the user becomes dynamic or the system may need to upgrade its product to remain up-to-date. However, the IoT application is subjected to increased security and privacy threats because an unauthorised user may be able to obtain access to highly delicate, consolidated business information \cite{alaqrabi2015}.

In a complex and challenging application there is a need to delegate access control mechanisms securely to one or more parties, who can in turn, control the methods that enable multiple other parties to authenticate with respect to the services they wish to consume. The primary challenge of any multiparty application is the need to authenticate customers so that they can be granted controlled access to information and data resources hosted on the cloud \cite{alaqrabi2018,alaqrabi2019}.

The wider distribution of of IoT nodes and the extent and nature of the data collected and transformed by such devices is a major challenge for security \cite{he2016}. In the IoT domain, authentication permits the integration of various IoT devices deployed in various contexts. In view of the fact that services and organisations can adopt a collaborative process in an extremely vibrant and flexible manner, direct cross-realm authentication relationship is not simply a means of joining the two collaborating realms.

The lack of authentication path connecting two security realms will necessitate two security realms \cite{hada2002}, when working together, to follow a more traditional and long route that will involve creating a mutual trust entailing entering into contractual agreements, multi-round cooperation and human intervention \cite{alaqrabi2018}.

%--------------------------------------------------------------------------------%
The primary reason for this lack of progress is due to serious concerns about the security, privacy, and reliability of these systems \cite{alaqrabi2013}. IoT is capable of monitoring all aspect of day-to-to life, including the above-mentioned concerns. Citizens, therefore, have legitimate concerns about privacy.

In addition, businesses are concerned with damage to their reputations due to data being handled by wrong hands, and the governments fear the consequences of security risks \cite{cao}. Multiparty authentication is a complex challenge in a multi-cloud environment.

These challenges increase in complexity when we consider the potential proliferation of devices in IoT systems. In general, such systems may be a one-to-one mapping between system access devices and the clouds themselves.

However, there are also several additional complications of numerous devices with varying degrees of functionality and capability. An example of such a device is a Wireless Sensor Network~(WSN), which are often adaptive entities that may be applicable to the addition or removal of sensor nodes during operation.

Various reports predict a remarkable increase in the number of connected intelligent `things' exceeding 20 billion by 2020 \cite{Ndiaye2017,Gartner2015}. As we see the exponential growth of the connected devices, the predictions seem to be believable. If these predictions come true, then the demand for authentication of devices will be a major challenge to address, especially as there will be insufficient capacity to manually authenticate even a fraction of the devices and consequently, some automation will be mandatory.

A fundamental challenge in a complex distributed computing environment that emerges as a consequence of the IoT and other combinations of technologies such as cloud architectures and microservices, is the necessity to manage and ensure that the required authentication approvals are in place to enable effective, secure communications.

For instance, Service Oriented Architectures (SOA) enable software systems to re-use `black box' functionality, by way of intra-service messaging that is facilitated by internet protocols. A more recent refinement of this is `microservices' architecture where there is a consideration of the level of granularity of service that is offered by software.

These paradigms, although abstract, offer considerable opportunities for system architects to develop resilient functionality within software, especially since re-used code can be comprehensively tested and secured. This approach to software development emphasises the need to develop secure systems, particularly since the IoT is in many cases a manifestation of a Cyber Physical System, whereby physical actuation is controlled and governed by software. Such systems present risks as well as opportunities, giving rise to the importance of formal approaches to the design of such systems.

As such, secure communication between software components is essential \cite{Modieginyan2017}, both in terms of ensuring that a particular message or instruction reaches the intended destination, but also that the complexity of out-sourced service functionality is provided with the appropriate authority to perform the task that is being requested.

%--------------------------------------------------------------------------------%
The use of Single Sign On~(SSO) \cite{clercg2002} also allows the use of a key exchange technique to actually manage the provision of authentication credentials certified by a named authority. In addition, it eliminates the need for users to enter different security credentials multiple times.

However, despite the relative simplicity of the technique, it simply provides a secure method of key exchange is insufficient for the situation when we need multiple parties to be capable to establish certain trust each other in a dynamic, heterogeneous environment \cite{xu2012}, and therefore SSO technique is lacking in this regard.

\subsection{A Multiparty Authentication model}

Prior work \cite{alaqrabi2018}, as described briefly below, describes a framework that addresses the challenges of obtaining the required authorisation agility in a dynamic multiparty environment.

This multiparty authentication model for dynamic authentication interactions is relevant when participants from various security realms want to access distributed operational data through a trusted manager. Al-Aqrabi et al \cite{alaqrabi2018} addressed issues related to reliable, timely and secure data transfer processes needed for shared company data processing networks.

This scenario is directly transferrable to the situation where large numbers of sensor nodes are producing streamed data, that requires real-time processing for the purposes of signal conditioning \cite{cao}, data cleansing, localised analytics processing, etc. For the sensor and computational nodes to work together in a service oriented way, there needs to be a mechanism where trusted access to data that is both in-transit and stored in a repository is feasible.

In addition, the multiparty authentication model can be used effectively to assist any distributed computing environment, for instance where cloud session users need to authenticate their session participants, and thus require a simplified authentication processes in multiparty sessions. 

Therefore, we have developed and extended this work to support the development of, for example, specific use cases where the availability of 5G network infrastructure can allow new business opportunities through enhanced performance. Figure 1 shows the framework for a \emph{Session Authority Cloud} that is applied as a certificate authority in this situation, although it could also be a remote cloud.

The $SAC$'s function is to control the individual sessions requested by any of the various parties (clouds). The $SAC$ does not differentiate between clouds and does not depend on them and it retains general authority over any party wishing to join the system. The $SAC$ retains authentication data for all tenants, including, for instance, root keys.

\subsection{An Authentication Protocol}

In this section, we introduce our proposed authentication protocol that addresses IoT application scenarios and data analytics applications that are accessed through IoT clouds, where participants from different security realms need to access distributed analytics services through a trusted principal.

This may apply especially when there are no direct authentication relationships in multiple IoT cloud systems between the people of different security realms and the distributed IoT cloud services.

Let $A$ be the trusted principal by which the requesting user approaches the $SAC$. The session authentication approval protocol starts with a user, $U$ who is a member of any security realm approved by the trusted principal. Providing access to IoT database objects in IoT clouds $A$ and $B$ is presumed, if the $SAC$ authorises the request forwarded by the principal. This is also presumed that $SAC$ will not accept any request that is not forwarded by the trusted principal.

The user who requests access is not a member of IoT $Cloud_A$ or a member of IoT $Cloud_B$. In principle, the user is a member of a security realm that is a different IoT cloud ($Cloud_C$) that the $SAC$ trusts.

The principal should know who the user is, since the $SAC$ mainly trusts the principal to accept the session request. $ID_r$ is the requesting user's cloud membership key. $ID_s$ is the requesting user's sub-domain membership and $ID_{sess}$ is the session key allocated by the $SAC$ to access IoT database files on IoT $Cloud_A$ and $Cloud_B$. IoT $Cloud_A$ and $Cloud_B$ will only open access to this key when authorised and forwarded by the $SAC$.

\subsection{Algorithm: Protocol for session approval}
\begin{algorithm}

 % \caption{Algorithm: Protocol for session approval}
  
  %\rule{8.8cm}{0.05cm}\\
  \textbf{Output}
  A value in variable $Flag$ to show that a session\\ is granted ($Flag = 1$) or denied ($Flag = 0$).\\
  \rule{12.2cm}{0.05cm}\\
  \textbf{Steps}
  \label{Algo:notexistingmember}
 % \begin{algorithmic}[1]
    %\textbf{Steps}
    \begin{enumerate}
 
\item $U_A$ to $F$ :request Access to IoT cloud 
\item $F$ to $U_A$ : request for the identity ID
\item $U_A$ to $F$ : $U_A$ sends the certificate $CA$ to $F$
\item $F$ to $SAC$ : session request sent to SAC 
\item $SAC$ to $SAC$-$DB$-$SH$: verifies $U_A$ identity

\item $SAC$ to IoT cloud: Flag indicating $U_A$ is authenticated\\
or not. Sends the SessionID and UserID to the\\ IoT
Cloud CA if authenticated.
\item IoT Cloud to SAC: stores the session ID and key in its \\ registry and then sends a reply 
\item $SAC$ to $F$ : sends a reply for session approval to
$F$\\ for authenticated user $U_A$.
\item F to $U_A$ : approves the decision to grand session\\ for authenticated user $U_A$. Flag = 1 and exit.
  \end{enumerate}
  %\rule{8.8cm}{0.05cm}
%\end{algorithmic}
\end{algorithm}

%The authentication protocol’s interactions are as follows:
%\item Are the keys matching?
%\item $Yes$/$No$
%\item $Proceeding with resources AuthoriseAccess$/$DenyAccess$
%\item $SAC$ to IoT Cloud :generates a session ID for the IoT cloud access.
%\item IoT Cloud to $SAC$
%\item $SAC$ to $F$
%\item $F$ to $UA$

%Proceeding with resources Access/Deny session
%\item Your session is approved – resources will be available soon/session denied due to wrong keys
%\item SAC to SAC session handler – Send access request for resources (with resources details and a session key) to cloud A and cloud B and inform requester
%\item Request for resources access (session key from SAC enclosed)
%\item Granted
%\item Request for resources access (session key from SAC enclosed)
%\item Session key and links to resources provided
%\item Session is established 
%
\begin{figure}[htbp]
\centering
\includegraphics[width=\textwidth]{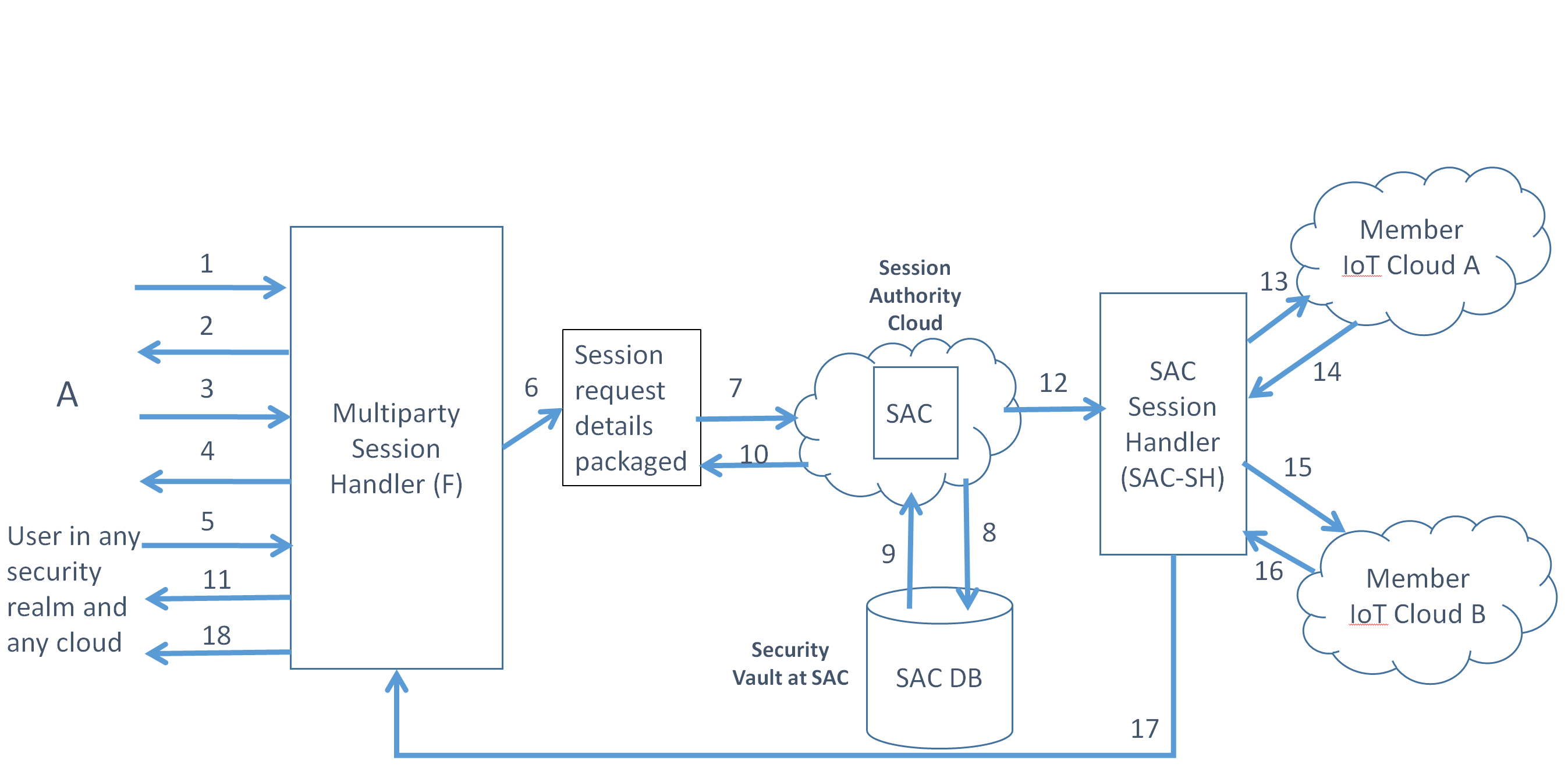}
\caption{Protocol for session approval.}
\label{fig:propmod}
\end{figure}

Figure \ref{fig:propmod} shows the session approval protocol. First user $U_A$ sends a request to create a new session in order to access IoT database objects in IoT $Cloud_A$ and $Cloud_B$. $F$ sends a request for $U_A$’s keys. $U_A$ sends his/her certificate, which contains a root key and subdomain key, and the certificate is encrypted with $U_A$’s private key. 
The multiparty session handler ($F$) generates a new session ID and sends it, along with $U_A$’s request. $SAC$ then verifies $U_A$ identity and to approve a new session. $SAC_DB$ uses $U_A$'s public key. $SAC$ also generates the key of the new session and then registers the session its session list.

Then it can verify $U_A$’s identity. If $U_A$’s identity is valid, then $SAC$ generates a session key and sends a request to access IoT clouds. After receiving a reply from resources, $SAC$ then sends a response of session approval with session key and available resources list.

IoT $Cloud_A$ stores the session ID and key in its registry and then sends a reply to $SAC$. $SAC$ sends a reply for session approval to $F$. Then, $F$ sends a response for session approval to access to access IoT database objects in IoT $Cloud_A$ and $Cloud_B$.

\section{Simulation approach}
The key focus of the work reported here was an in-depth exploration of how the reduction in connection latency offered by 5G systems, in comparison to 4G systems, impacts the performance of our novel multiparty authentication protocol.  

\begin{figure}[htbp]
\centering
\includegraphics[width=0.85\textwidth]{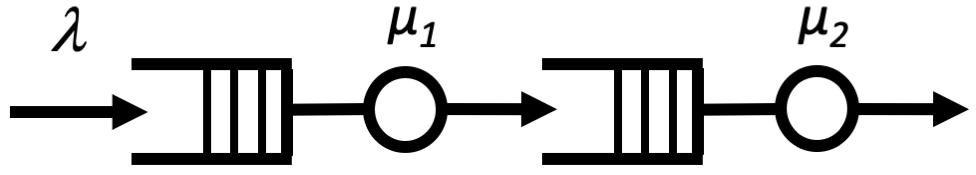}
\caption{Cascaded M/M/1 queue.}
\label{fig:casc_q}
\end{figure}

The simulation was built to enable the comparison of the delay experienced by a party seeking authorisation as the rate of authentication requests varied.  The authentication delay is conditioned by two factors:
\begin{itemize}
\item The time taken for the authentication request to be transported by the mobile part of the system - there are two aspects to this parameter: the time taken to transmit the bits; and other sources of latency in the mobile system such as scheduling, resource allocation and routing
\item The time taken for the authentication request to be processed by the authentication server.  Earlier simulation work suggest that the time to service a request with reasonable hardware is of the order of 6~ms \cite{alaqrabi2018}.
\end{itemize}
When reduced to its fundamental structure, the entire mobile radio system and authentication server systems can be modelled as two cascaded queues as shown in Figure \ref{fig:casc_q}.  Authentication requests are generated by a pool of devices with a Poisson distribution wholly defined by a generation rate.  

These requests are queued and the service rate of this first M/M/1 queue is dependent on a combination of the mobile system latency mentioned above and the data-rate dependent time taken to transmit the authentication request package.  A typical packet size is of the order of 1~kbit, and the transmission delay associated with transporting this is of the order of 0.2~ms at 5 Mbit/s, or 0.02~ms at 50~Mbit/s.

Compared to the overall latency of the end-to-end transmission, and typical service times of the authentication server, these times are negligible and are safely ignored for the rest of the simulation with the service rate of the queue modelling the radio part of the system being determined solely by a single parameter with the system data-rate being irrelevant in this context.

The authentication requests are then passed from the queue modelling the radio part of the system to a second queue which models the latency of the granting of the authentication request by setting an appropriate service rate for that queue.
The implementation of such a system requires the utilization of an event-driven simulation approach, and in Python, the SimPy \cite{SimPy} package was selected as the framework for the simulation as, when combined with SimComponents.py \cite{SimComp}, queue models can be assembled and executed with relative ease, and comprehensive performance data is readily accessible for further analysis.
\section{Results}
\begin{figure}[htbp]
\centering
\includegraphics[width=0.9\textwidth]{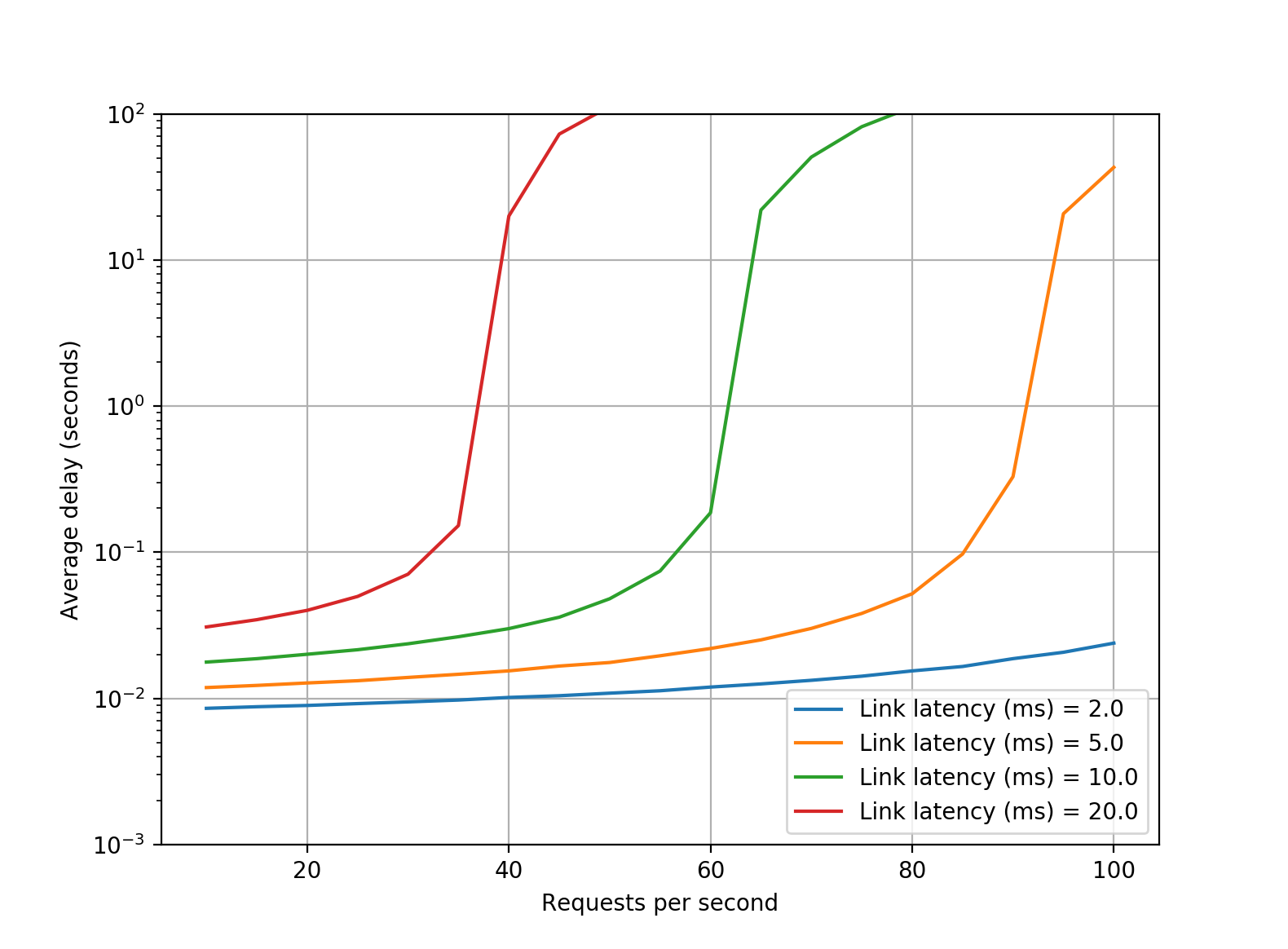}
\caption{Average (mean) authentication delay for a 6~ms authentication request service time}
\label{fig:6ms}
\end{figure}

\begin{figure}[htbp]
\centering
\includegraphics[width=0.95\textwidth]{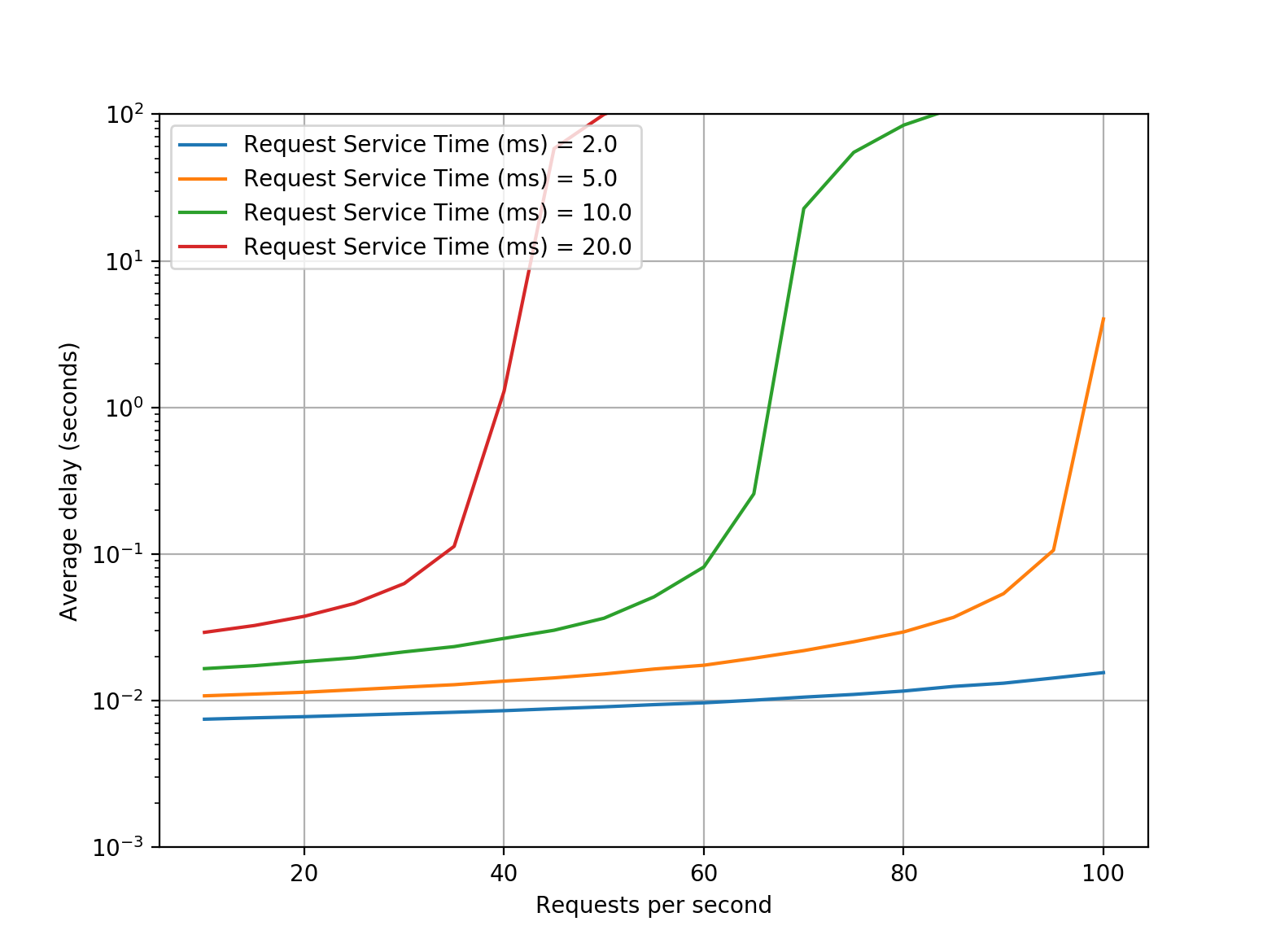}
\caption{Average (mean) authentication delay for a link latency of 50~ms}
\label{fig:5ms}
\end{figure}

\begin{figure}[htbp]
\centering
\includegraphics[width=0.95\textwidth]{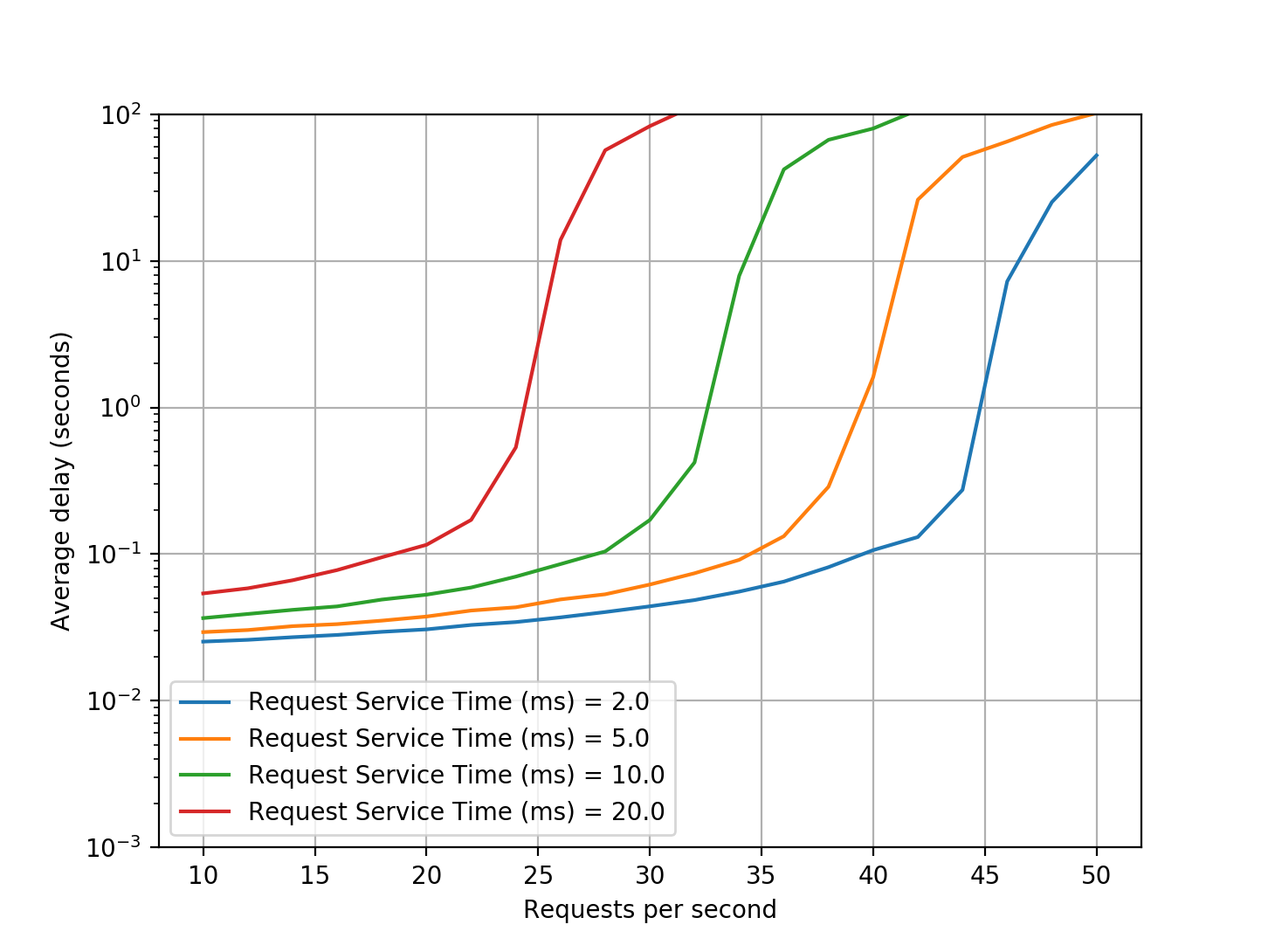}
\caption{Average (mean) authentication delay for a link latency of 20~ms}
\label{fig:20ms}
\end{figure}

Figure \ref{fig:6ms} shows how the average (mean) delay experienced by an authentication request varies with the rate of requests for a number of different link latencies.  The results in this figure are based on the 6~ms authentication delay discussed above.  As expected, a higher request rate and/or greater link latency yields a greater average authentication delay.
Figure \ref{fig:5ms} fixes the link latency at 5~ms -- a typical, conservative, value for a 5G system -- and explores how average authentication delay varies with request rate and the authentication request service time.  Figure \ref{fig:20ms} presents the same information, but this time with an assumption of a 20~ms link latency -- a value that is typical of a well performing 4G system.  Again, both of these results follow the expected form.

\begin{figure}[htbp]
\centering
\includegraphics[width=0.95\textwidth]{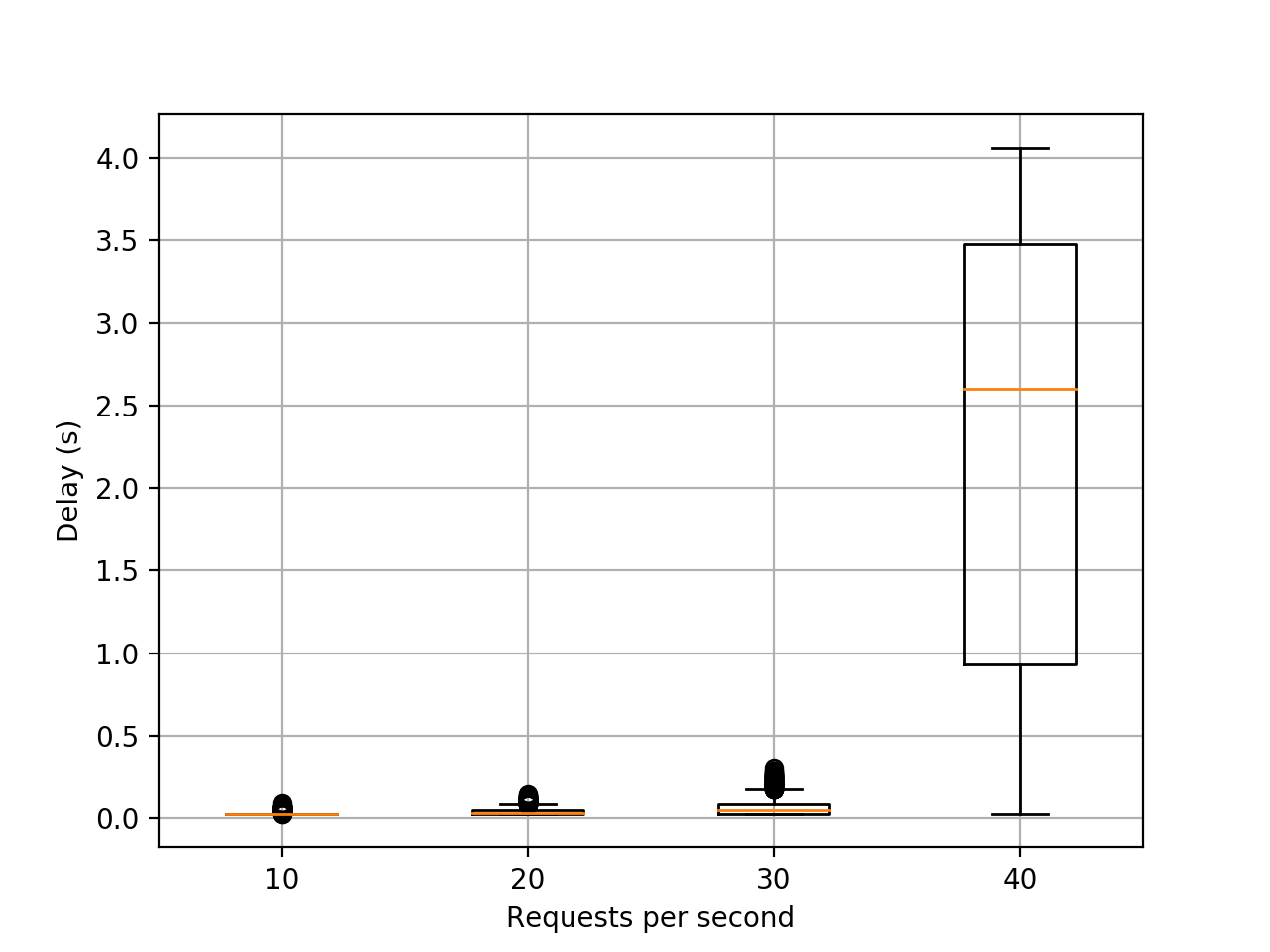}
\caption{Distribution of authentication delays for a 20~ms link latency}
\label{fig:bp20}
\end{figure}

\begin{figure}[htbp]
\centering
\includegraphics[width=0.95\textwidth]{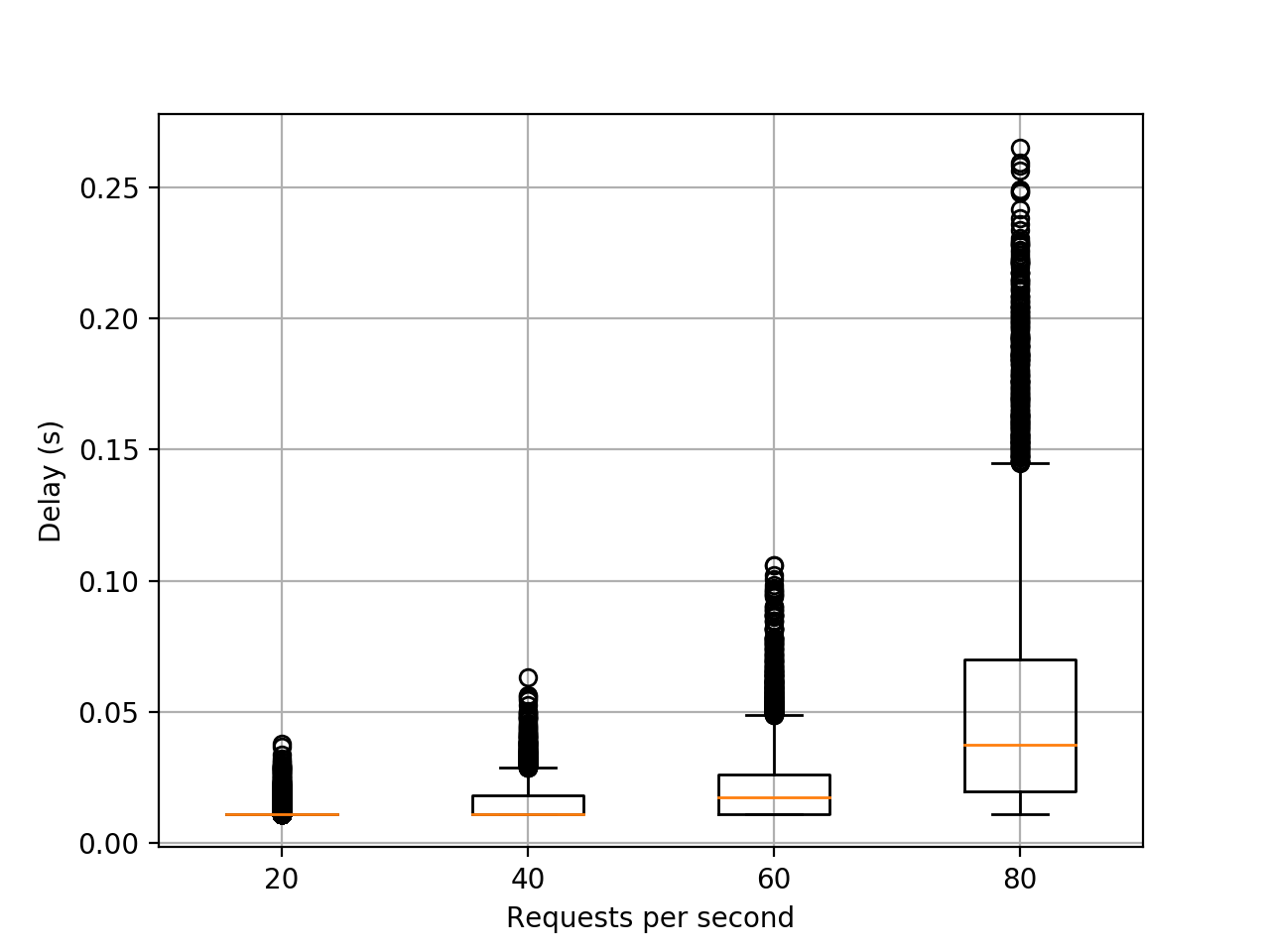}
\caption{Distribution of authentication delays for a 5~ms link latency}
\label{fig:bp5}
\end{figure}

The previous three figures only look at the mean delay for an authentication request.  In a real-world deployment, we would often have considerable interest in the range and distributions of the actual delays experienced by individual authentication requests.  Figures \ref{fig:bp20} and \ref{fig:bp5} show the distribution of authentication requests as box-plots for a range of request rates.  Figure \ref{fig:20ms} is for a 20~ms link latency, and Figure \ref{fig:bp5} is for a link latency of 5~ms.  The whiskers adopt the customary convention of defining points that fall outside of $\pm 1.5 \times IQR$ either side of the median as outliers.

\section{Discussion of results}

For most application scenarios, authentication requests will need to be provided in a relatively timely manner.  If, for example, we impose an upper limit of 0.1~s for an authentication request to be serviced, then Figure \ref{fig:6ms} shows that for a link latency typical of a 4G system -- 20~ms -- then only slightly over 30 requests per second can be accommodated. 

This in contrast to the situation when the link latency is lower.  With a latency of 2~ms, 100 requests per second can be accommodated with only $\approx$~0.02~s of delay, and even with a link latency of 5~ms, 85 requests per second can be accommodated within a 0.1~sec mean delay. These results show that the adoption of 5G is a key enabler of the enhanced multiparty authentication mechanism described in this research, as the link performance of 4G systems severely inhibits the usefulness of the authentication approach.

Figure \ref{fig:5ms} considers a 5G-like scenario with a 5~ms link latency and considers the impact of the time required to service an authentication request.  This part of the investigation can help to inform the dimensioning of the authentication server hardware.  Taking the same 0.1~s target for mean authentication delay, it can be seen that a service request time of 20~ms severely limits the overall system performance with only some 35 requests per second being accommodated.

Conversely, a service request delay of 10~ms yields an ability to support some 95 requests per second.  In contrast, Figure \ref{fig:20ms} shows that for a 4G-like scenario with a 20~ms link delay, the benefits achieved by improving the performance of the authentication server are limited.  Even with a 5~ms service time, only some 35 requests per second can be accommodated and importantly, the inherent latency of the radio link limits the overall throughput even if authentication server response times are pushed to levels that are not easily achievable in practice.

The lower latency offered by 5G amplifies the benefits offered by improving server performance, and excellent overall throughput can be achieved with server response times that are reasonably realisable.  

Mean delays are only one part of the overall system performance assessment as they only allow us to quantify the service as experienced across the user base on an aggregated basis.  The spread of delays experienced by devices attempting to authenticate are also of interest, as for a particular node attempting to authenticate, it is the delay experienced by that one node that is of importance. Figure \ref{fig:bp20} shows that for a 4G-like system with a link latency of 20~ms and an authentication response time of 6~ms, that even with 30 requests per second (where the mean response time was reasonable at under 0.1~s), a considerable number of requests had delays of significantly more than this, up to around 0.4~s as a worst case.

Figure \ref{fig:bp5} shows that for a 5G-like system then nearly all of the requests at a rate of 60 requests per second achieve an authentication time under 0.1~s, and at 80 requests per second, 75~\% of the requests are handled within 0.08~s, and all but a very few within 0.25~s. Again, this emphasises the key enabling role that 5G deployment will play in facilitating the deployment of real-world systems based on the enhanced authentication protocol described here.
\section{Conclusions}

This research explores issues around the authentication of large numbers of devices in an Iot or IIoT scenario.  We describe some of the challenges that emerge as (I)IoT systems grow in scale, and consider how the improved network performance offered by 5G particularly in terms of latency opens up opportunities to deploy robust, flexible, dynamic authentication protocols that can improve the trust placed in (I)IoT devices, and thereby facilitate their use in situations where business critical intellectual property (IP) could be exposed if security were inadequate.

We then proceed to describe a dynamic and flexible authentication protocol and explore how this performs under constraints of latency that would be experienced in both 4G and 5G networks.  Our findings show clearly that the reduced latency offered by 5G mobile systems is a critical enabler of the delivery of overall system performance that makes our authentication protocol a realistically deployable option with a performance level that would enable its use in high density, dynamic IIoT applications.


\begin{thebibliography}{8}
%
\bibitem{kalra2015}
S. Kalra and S. K. Sood. Secure authentication scheme for IoT and cloud servers. Pervasive and Mobile Computing, Elsevier, 24, 2015, pp. 210-223.
%
\bibitem{alaqrabi2015}
H. Al-Aqrabi, L. Liu, R.Hill, N. Antonopoulos. Cloud BI: Future of business intelligence in the Cloud, Journal of Computer System Science, Elsevier, 2015.
%
\bibitem{ferrag2018}
M. A. Ferrag, L. Maglaras, A. Argyriou, D. Kosmanos, and H. Janicke,
“Security for 4G and 5G cellular networks: A survey of existing authentication
and privacy-preserving schemes,” J. Netw. Comput. Appl.,
vol. 101, pp. 55-82, Jan. 2018.
%
\bibitem{ldxu2014}
L.D. Xu, W. He, S. Li, Internet of things in industries: a survey, IEEE Trans. Ind. Inf.
10 (4) (2014) 2233–2243.
%
\bibitem{hosek2013}
I.J. Hosek, Enabling technologies and user perception with integrated 5g-iot ecosystem,
2016.
%
\bibitem{Ishaq2013}
I. Ishaq, et al., IETF Standardization in the field of the internet of things (iot): a
survey, J. Sensor Actuator Netw. 2 (2) (2013) 235–287.

%
\bibitem{Akpakwu2017}
G.A. Akpakwu, et al., A survey on 5g networks for the internet of things: communication
technologies and challenges, IEEE Access (2017).
%


\bibitem{Elkhodr2017}
M. Elkhodr, S. Shahrestani, H. Cheung, The internet of things: new interoperability, management and security challenges, 2016, Arxiv:1604.04824.
%
\bibitem{alaqrabi2013}
H. Al-Aqrabi, L. Liu, R. Hill, N. Antonopoulos, A Multi-layer Hierarchical Inter-Cloud Connectivity Model for Sequential Packet Inspection of Tenant Sessions Accessing BI as a Service. Proceedings of 6th International Symposium on Cyberspace Safety and Security and IEEE 11th International Conference on Embedded Software and Systems. France, Paris, March 20-22, IEEE, 2014 pp. 137-144.
%



\bibitem{he2012}
He, D., 2012. An efficient remote user authentication and key agreement protocol formobile client and server environment from pairings. Ad Hoc Netw. 10 (6), 1009–1016.
%
\bibitem{deng2009}
Deng, Yaping, Fu, Hong, Xie, Xianzhong, Zhou, Jihua, Zhang, Yucheng, Shi, Jinling,
2009. A novel 3GPP SAE authentication and key agreement protocol. In: Proceedings of International Conference Networks Infrastructure and Digital Content, IEEE, pp.
557–561.
%
\bibitem{kar2011}
Karopoulos, G., Kambourakis, G., Gritzalis, S., 2011. PrivaSIP: ad-hoc identity privacy in
SIP. Comput. Stand. Interfaces 33 (3), 301–314.
%
\bibitem{Alrawais2017}
A. Alrawais, A. Alhothaily, C. Hu, X. Cheng. Fog computing for the Internet of Things: Security and privacy issues, IEEE Internet Comput., 21(2), pp34-42, Mar 2017.
%
\bibitem{barni2010}
M. Barni et al., Privacy-preserving fingercode authentication. In: Proceedings The 12th ACM Workshop on Multimedia and Security - MM Sec ’10, ACM Press, New York, New York, USA, p. 231.
%
\bibitem{ma2014}
Ma, C.-G., Wang, D., Zhao, S.-D., 2014. Security flaws in two improved remote user authentication schemes using smart cards. Int. J. Commun. Syst. 27 (10),
%
\bibitem{mahmoud2016}
Mahmoud, M., Saputro, N., Akula, P., Akkaya, K., 2016. Privacy-preserving power injection over a hybrid AMI/LTE smart grid network. IEEE Internet Things J., 1.
%
\bibitem{hill2017}
D. Shadija, M. Rezai, R. Hill.  Towards an Understanding of Microservices. Proceedings of the 23rd International Conference on Automation \& Computing, University of Huddersfield, 7-8 September, 2017, IEEE.
%
\bibitem{bessis2013}
N. Bessis, F. Xhafa, D. Varvarigou, R. Hill, M. Li, (Eds.) Internet of
Things and Inter-cooperative Computational Technologies for Collective
Intelligence. In Studies in Computational Intelligence, Springer,
2013.
%
\bibitem{hill2013}
R. Hill, J. Devitt, A. Anjum, M. Ali. Towards In-Transit Analytics
for Industry 4.0. IEEE International Conference on Internet of Things
(iThings) and IEEE Green Computing and Communications (Green-
Com) and IEEE Cyber, Physical and Social Computing (CPSCom)
and IEEE Smart Data (SmartData). IEEE Computer Society.
%
\bibitem{sotiriadis2013}
S. Sotiriadis, N. Bessis, N. Antonopoulos, R. Hill. Meta-scheduling algortithms for managing inter-cloud interoperability. International Journal of High Performance Computing and Networking, 7(2), pp156--172, 2013.
%
\bibitem{baker2012}
C. Baker, A. Anjum, R. Hill, N. Bessis, S. Liaquat Kiani. Improving cloud datacentre scalability, agility and performance using OpenFlow, Proceedings of the 4th International Conference on Intelligent Networking and Collaborative Systems (INCoS), pp1--15, 2012, IEEE.
%
\bibitem{alaqrabi2018}
H. Al-Aqrabi, R. Hill. Dynamic Multiparty Authentication of Data Analytics Services within Cloud Environments. In 20th IEEE International Conference on High Performance Computing and Communications (HPCC-2018), Exeter 28-30, IEEE.
%

\bibitem{alaqrabi2019}
H. Al-Aqrabi, R. Hill. Dynamic Multiparty Authentication of Data Analytics Services within Cloud Environments. Proceedings of the 20th International Conference on High Performance Computing and Communications, 16th International Conference on Smart City and 4th International Conference on Data Science and Systems, HPCC/SmartCity/DSS2018. IEEE Computer Society, pp742--749, 2018
%

\bibitem{he2016}
D. He, S. Zeadally, L. Wu, and H. Wang. Analysis of handover authentication protocols for mobile wireless networks using identity-based public key cryptography, Comput. Networks, Dec 2016.
%
\bibitem{hada2002}
S. Hada, H. Maruyama. Session Authentication Protocol for Web Services, Proc. Symposium on Application and the Internet, pp158-165, 2002.
%
\bibitem{cao}
N. Cao, S.B. Nasir, R.A. Shreyas Sen, Self-optimizing IoT wireless video sensor node with in-situ data analytics and context-driven energy-aware real-time adaptation,
IEEE Trans. Circuits Syst. I Regul. Pap. 64(9), pp2470--2480, 2017.
%
\bibitem{Ndiaye2017}
M. Ndiaye, G.P. Hancke, A.M. Abu-Mahfouz, Software defined networking for improved
wireless sensor network management: a survey, Sensors 17 (5) (2017) 1–32.
%
\bibitem{Gartner2015}
Gartner. ``Gartner Says 6.4 Billion Connected Things Will Be in Use in 2016, Up 30 Percent From 2015?'', Gartner website, http://www.gartner.com/newsroom/id/3165317, 10th November, 2015.
%
\bibitem{Modieginyan2017}
K.M. Modieginyane, B.B. Letswamotse, R. Malekian, A.M. Abu-Mahfouz. Software defined wireless sensor networks: application opportunities for efficient network
management: a survey, Comput. Electr. Eng. 1--14, 2017.
%
\bibitem{clercg2002}
J. D. Clercq. Single Sign-On Architectures, Proc. International Conference, InfraSec 2002, Bristol, UK, pp40--58, 2002.
%
\bibitem{xu2012}
J. Xu, D. Zhang, L. Liu, X. Li, X. Dynamic Authentication for Cross-Realm SOA-Based Business Processes, IEEE Transactions on services computing, 5(1), pp20--32, 2012.
%


%
\bibitem{SimPy}
SimPy documentation. https://simpy.readthedocs.io/en/latest/index.html. Accessed October 2019.
%
\bibitem{SimComp}
Grotto Networking, Basic Network Simulations and Beyond in Python. https://www.grotto-networking.com/DiscreteEventPython.html. Accessed August 2019.
%



\end{thebibliography}
\end{document}